\documentclass[12pt,preprint]{aastex}

\slugcomment{Accepted by The Astrophysical Journal}

\shorttitle{Impey et al.}
\shortauthors{The Binary Quasar LBQS~0015+0239}

\begin{document}


\title{LBQS~0015+0239: A Binary Quasar with Small Angular 
Separation\altaffilmark{1}}
\altaffiltext{1}{The observations reported here were obtained at the
W. M. Keck Observatory, which is operated as a scientific partnership among
the California Institute of Technology, the University of California, and
the National Aeronautics and Space Administration. The observatory was made
possible by the generous financial support of the W. M. Keck Foundation.}


\author{Chris D. Impey and Cathy E. Petry}
\affil{Steward Observatory, University of Arizona, Tucson, AZ 85721}
\email{cimpey@as.arizona.edu, cpetry@as.arizona.edu}

\author{Craig B. Foltz}
\affil{Multiple Mirror Telescope Observatory, University of Arizona, 
	Tucson, AZ 85721}
\email{cfoltz@as.arizona.edu}

\author{Paul C. Hewett}
\affil{Institute of Astronomy, Madingley Road, Cambridge CB3 0HA, UK}
\email{phewett@ast.cam.ac.uk}

\and

\author{Frederic H. Chaffee}
\affil{W. M. Keck Observatory, 65-1120 Mamalahoa Highway, Kamuela, HI 96743}
\email{fchaffee@keck.hawaii.edu}


\begin{abstract}

We present spectroscopic observations with Keck/LRIS of LBQS~0015+0239, a 
pair of quasars at $z = 2.45$ with a separation of $\Delta\theta = 2.2$ 
arcsec (projected linear distance of 17.8 $h_{70}^{-1}$ kpc, for $\Omega_m 
= 0.3$, $\Lambda = 0.7$). Lensing is an unlikely interpretation for the 
images, since the spectra show significant differences in the \ion{N}{5} and 
\ion{C}{4} emission line profiles, and there is no luminous galaxy at the 
anticipated lens position. Rather, we interpret this pair as the highest 
redshift known example of a binary quasar. The redshift difference of
$661\pm173$ kms$^{-1}$ between the two components is inconsistent with 
the lensing expectation, but is consistent with the line of sight velocity 
difference of a bound pair of galaxies. The spectra show associated, narrow
metal absorption in highly ionized species of C and N, with a systemic 
velocity difference of $492\pm6$ kms$^{-1}$. It is likely that this absorption 
arises in a group or cluster of galaxies surrounding the quasar pair. After a 
thorough search, the Large Bright Quasar Survey (LBQS) is known to contain 
one gravitational lens and four probable binary pairs. The existence of 
close binaries is evidence that quasar activity can be triggered by galaxy 
interactions. We speculate that the close pair LBQS~0015+0239 is a rare 
example of the pre-merger state of two supermassive black holes.

\end{abstract}




\section{Introduction}

The rare situations of closely-separated quasars at the same redshift 
are of particular interest for observational cosmology. The paired
images occur at a rate of $\sim 1$ in 500 quasars, and they are either 
the result of gravitational lensing or of physically distinct quasars 
in a binary system. Many of the pairs (and all of the triples and higher
multiplicities) are the result of gravitational lensing by foreground 
mass distributions. Over 60 lens systems are known (see the review of
Surdej \& Claeskens 2002, and the constantly updated list at the CASTLES 
web site {\url http://cfa-www.harvard.edu/castles/}). 
To be confirmed as a lens, 
the primary requirements are identical quasar redshifts, and a luminous
galaxy at a plausible position and redshift to cause the observed image
splitting and brightness ratio (Blandford \& Narayan 1992; Schneider,
Ehlers, \& Falco 1992; Kochanek 1995). Almost all of these systems have
separations smaller than $\Delta\theta = 3$ arcsec. At larger separations, 
there is a set of about 20 quasar pairs where the interpretation is 
not secure. A system is best confirmed as a binary if the spectral 
energy distributions are substantially different, including, for 
example, cases where one quasar in the pair is radio-loud and the 
other is radio-quiet. The binary population is important because it 
leads to the inference that quasar activity can be fuelled by galactic 
interactions and mergers (Djorgovski 1991; Peng et al. 1999; Kochanek,
Falco, \& Mu\~noz 1999; Mortlock, Webster, \& Francis 1999).

It is difficult to decide with any certainty whether an individual
pair is a lens or a binary. The presence of a luminous, and usually 
early type, galaxy along the line joining the two images is a good
indication of lensing, since the chance probability of such alignment
is very small. However, in some cases the available imaging is not deep
enough to detect the putative lensing galaxy at all plausible redshifts,
and the possibility of ``dark'' galaxies cannot be entirely ruled out
(Hawkins 1997). Image separations of $\Delta\theta =$ 1 to 3 arcsec are 
indicative of lensing but not prescriptive, since binaries in a final 
pre-merger state might have small projected separations. Conversely, 
the larger separations of $\Delta\theta =$ 3 to 10 arcsec need not be 
binaries, since galaxy lensing can get an assist from a surrounding 
cluster potential (Walsh, Carswell, \& Weymann 1979; Lawrence et al. 
1984; Wisotzki et al. 1993). Measurement of out-of-phase but identical 
fractional flux variations between the two images would be sufficient 
evidence to declare a lens; but this type of data has not been acquired 
for any system where the lensing interpretation was not already secure.
Another prediction of lensing is achromaticity. However, a clean test 
for spectral differences is complicated by the generally high degree 
of similarity of quasar spectra (Francis et al. 1992), the intrinsic
variations in quasar continua and broad lines (Small, Sargent, \&
Steidel 1997), and the possibility of differential extinction and
reddening from a dusty lens (Falco et al. 1999). Spectral differences
can also be caused by microlensing, where a stellar mass object can
temporarily magnify the continuum of one image relative to the 
emission lines (Wisotzki et al. 1993; Lewis et al. 1998).

The status of an individual pair can only be evaluated from the ensemble 
of its observed properties. Several recent papers have attempted to do 
this. Kochanek, Falco, \& Mu\~noz (1999) and Peng et al. (1999) both 
worked primarily with data gathered as part of the CfA/Arizona Space 
Telescope Lens Survey (CASTLES), and Mortlock, Webster, \& Francis (1999) 
applied their own criteria to cataloged quasar pairs. These lists are in 
broad agreement, containing 13, 14, and 16 binaries respectively, after 
subtracting off the pairs where lenses have subsequently been detected. 
Two out of the 16 binaries listed by Mortlock et al. are noted as being 
possibly lensed. With Q~1208+1011, the luminosity and small separation 
are suggestive of lensing. Recent Hubble Space Telescope observations 
have failed to detect a lens, but the small separation and high contrast 
does not allow a sensitive limit (L\'ehar et al. 2000). Since there is 
no actual evidence that it is a binary, it is preferable to consider it 
as a probable lens. In the case of the wider separation pair Q~2345+007, 
lensing was intially indicated by the strikingly similar spectra (Steidel 
\& Sargent 1991). However, recent Chandra observations find differences 
in the overall energy distributions that strongly indicate that this 
system is a binary (Green et al. 2002). The number of possible binaries 
increases from 15 to 19 if the objects discovered subsequent to the 
Mortlock et al. paper are included: CLASS B0827+525 (Koopmans et al. 
2000), CTQ 839 (Morgan et al. 2000), the serendipitously discovered 
LBQS~0103--2753 (Junkkarinen et al. 2001), and LBQS~0015+0239 (this 
paper).

Another approach is to compare the statistical properties of the
population of possible binaries to the properties of confirmed lens
systems. The main caveat to this approach is the inhomogeneous nature 
of the sample of known pairs, which come from a wide variety of radio 
and optical surveys. The detection of one or both quasars as radio sources 
provides additional discriminatory power, since the radio-to-optical flux 
ratios of quasars vary over a wide range, and most quasars are radio-quiet 
(Hooper et al. 1996; Bischof \& Becker 1997). Kochanek et al. (1999) have 
used the absence of pairs where both quasars are radio-loud, and the 
incidence of pairs where one quasar is radio-loud, to argue that 
essentially all the pairs where no lens is seen are in fact binaries. 
In addition, the distribution of redshifts, flux ratios and separations 
of the pairs is different from comparable distributions for confirmed 
lenses (Mortlock et al. 1999). The spectral similarity of the quasar 
pairs is suggestive, but not compelling, evidence in the other 
direction (Peng et al. 1999; Mortlock et al. 1999).

In this paper, we present observations of a small separation quasar 
pair that we interpret as a probable binary system. The new system is 
noteworthy because of the high redshift, the small angular separation, 
and the fact that it comes from a well-studied quasar catalog. Although 
the Large Bright Quasar Survey (LBQS) has been eclipsed in size by the 
2dF and SDSS quasar surveys (Richards et al. 2001; Croom et al. 2001), 
it has been the subject of a large amount of follow-up activity, and 
its completeness has been calibrated with respect to the FIRST survey 
(Hewett, Foltz \& Chaffee 2001). Most importantly, it has been thoroughly
searched for small separation pairs (Hewett et al. 1998). In section 
2, we present Keck/LRIS observations which resolve the components of 
this 2.2 arcsec separation pair. The spectra are noteworthy for the 
associated heavy element absorption that is seen in each component. In 
section 3, we present the evidence that LBQS~0015+0239 is a binary 
rather than a lens. Section 4 contains comments  on the incidence of 
lenses and binaries in the LBQS, and section 5 offers brief comments 
on the overall population of binary quasars. Conclusions follow in 
section 6.

\section{Observations}

All the LBQS quasars have been inspected visually on UK Schmidt
Telescope plate material to provide a census of companion objects
within a radius of $10\farcs0$ (Hewett et al. 1998). Companions with
magnitudes brighter than $B_J=21.5-22$, depending on the quality and
depth of the $B_J$ plates, have been identified. However, in practice,
the physical extent of the high density images of the relatively bright
quasars on the plates precludes the detection of faint companions closer 
than $\sim 3\farcs0$ and LBQS~0015+0239 appeared as a single object on 
the original plate material. A supplementary program of broad band imaging 
at optical wavelengths, using a variety of telescopes, has produced CCD 
images of 230 LBQS quasars. The majority of the images were obtained 
under conditions of good seeing ($\le 1\farcs0$), but are not particularly
deep. The companion to LBQS~0015+0239 was identified in $VRI$ images
obtained in 1996 at the 1m Jacobus Kapteyn Telescope and the image 
pair was flagged as a potential gravitational lens/binary quasar.

The quasar pair LBQS~0015+0239 was observed on the night of 2001 July 21 
with the 10m Keck I telescope and the Low Resolution Imaging Spectrograph 
(Rodgers \& McCarthy 1994). Observations were made in ``Red Only'' mode.  
The Red CCD is a 2048x2048 SITe/Tektronix with 24$\mu$m pixels, giving 
0.215 arcsec/pix averaged over the field of view. The 600 l/mm grating 
was centered just off its 5000\AA\ blaze at 5310\AA\ providing wavelength
coverage of 4000-6621\AA\, with 1.28 A/pix dispersion and a resolution 
of 4.2\AA. The slit width was 1 arcsec and the slit was aligned with the 
two images. Three 15 minute exposures were obtained in approximately 0.7
arcsec seeing conditions, which was sufficient to resolve the two images. 
Flatfields were taken, a flux standard was observed (BD+28$^\circ$4211), 
and HgNeAr and CdZn arc lamps were obtained for both the science and 
calibration frames.

The data was reduced using standard IRAF\footnote[2]{The Image Reduction 
and Analysis Facilities package is distributed by NOAO, which is operated 
by AURA Inc. under contract to the National Science Foundation.} routines.
The dual amp readout of the LRIS CCD necessitates the use of the Keck 
Observatory package {\it wmkolris} for overscan and bias subtraction. 
The flatfielding required an extra step in the determination of the CCD 
response function because of the discontinuity introduced by the dual 
amp readout of the chip. Because the standard routines do not allow for 
discontinuities, we simply fit each half of the chip independently using 
the standard routines, then joined the two halves to obtain the complete 
response function. Before extracting the one-dimensional spectra, the 
three frames are shifted and median combined to eliminate cosmic rays.
A variance weighted extraction is performed for each quasar component, 
including flux from $\pm$ 4 pixels to either side of the peak of the 
spatial profile. To obtain a summed spatial profile, the final combined 
image is corrected for geometrical distortion and pixels corresponding 
to $\lambda\lambda$~4600-6400 are summed along the dispersion axis using 
the IRAF task {\it blkavg}. All of the major absorption features seen in 
the combined spectrum are present in the individual 15 minute exposures.

Spectra of the two components of LBQS~0015+0239 are shown in the panels
of Figure~1. Prominent emission lines are labelled, and narrow absorption
lines are marked along the continuum. The selection of absorption lines is
based on software that uses a similar detection strategy to that used by
the Quasar Absorption Line Key Project (Schneider et al. 1993). The basic
features of the software are given in Petry, Impey, \& Foltz (1998), and
additional details can be found in Petry et al. (2002). The continuum is
modelled with a spline fit that adequately follows the rapidly changing 
flux in the vicinity of the broad emission lines. Absorption features are 
selected using an interative technique that assumes unresolved lines and
can deblend multiple Gaussian profiles in regions of complex absorption.

Tables 1 and 2 list the significant absorption lines detected in the two 
components A and B of LBQS~0015+0239.
Significance is defined in two ways: $SL_{fit}$ which reflects how well the 
absorber is fit by the software, and $SL_{det}$ which refers to the detection 
limit of the data.  $SL_{fit}$ is simply the ratio of the fitted equivalent 
width to its error, $W/\sigma_{fit}$.  $SL_{det}$ is different in that its
denominator, $\sigma_{det}$, is the $1\sigma$ limiting equivalent width, 
which is a function of wavelength and is computed using the flux error.
To obtain a reliable list of absorbers, we use a joint criterion for 
declaring significant lines: $SL_{fit} \geq 3$ and $SL_{det} \geq 4$.  
Additionally, some lines in blended regions can have very low $SL_{fit}$ 
even though the detection limit of the data may be low, resulting in a 
higher $SL_{det}$. Therefore, we include any line with $SL_{det} \geq 6$.
The RMS error in the wavelength calibration is 0.07\AA, although the absolute
error is higher shortward of 4500\AA, where there are very few arc lines.
Differential wavelength errors are smaller, but in practice the precision 
of measuring wavelength differences between the spectra is limited by the
error on the fitted central wavelength of each feature, as can be seen in
Table 1.

The difference in signal-to-noise ratio between the two spectra reflects 
the fact that component A is nearly ten times brighter than component B. 
Each spectrum shows strong Ly$\alpha$, \ion{S}{4}, and \ion{C}{4} emission, 
with the red end of the Ly$\alpha$ forest seen below 4200\AA. \ion{N}{5} 
is strong in the wings of Ly$\alpha$ in component A and virtually absent 
in component B. Strong and narrow associated absorption lines are seen 
eating into the Ly$\alpha$, \ion{N}{5}, \ion{Si}{4}, and \ion{C}{4} 
emission features. The redshifts of the 
quasars were measured by cross-correlation with the SDSS composite, using 
the IRAF routine {\it fxcor}. The results are $z = 2.4747\pm0.0043$ for 
component A and $z = 2.4831\pm0.0044$ for component B. The difference of 
$\Delta v = 720\pm380$ kms$^{-1}$ differs from zero at the $2\sigma$ level. 
Redshifts from cross-correlation are reliable because they use the full 
information in the spectra. However, the two components of LBQS~0015+0239 
are more similar to each other than they are to the SDSS composite (as 
indicated by a higher correlation amplitude), so the most precise measure 
of the velocity difference comes from the cross-correlation between A and 
B. The cross-correlation result is $\Delta v = 661\pm173$ kms$^{-1}$, which
is formally different from zero at the $4\sigma$ level. The amplitude
of the cross-correlation is 0.92, indicating the high degree of spectral 
similarity of the two components.

The major emission lines in each of the two components are stronger, by factors
of 1.4 (Ly$\alpha$/\ion{N}{5}), 2.3 (\ion{Si}{4}/\ion{O}{4}), 2.0 (\ion{C}{4}), 
and 3.5 (\ion{He}{2}) than the rest equivalent widths for the SDSS composite 
spectrum (Vanden Berk et al. 2001). The agreement between the two components 
is in all cases better than 35\%. Peng et al. (1999) found that the spectral 
similarities of the binary pairs were significantly greater than would be 
expected for an equivalent sample of randomly selected quasars (but see also 
Mortlock et al., who reach a different conclusion for two of the LBQS pairs). 
The spectra of the two components of LBQS~0015+0239 raise again the question 
of how two quasars with an order of magnitude different luminosity could
have such similar properties on parsec scales. 

The associated absorbers have a systematic velocity difference of $492\pm6$ 
kms$^{-1}$ between the two components. This number is based on a velocity
error weighted combination of the offsets between A and B of four different
species, listed in Table~3.  The two components of \ion{C}{4} are combined
into a single error-weighted offset before averaging with the other three
species.  Table~1 also lists Ly$\alpha$
absorbers seen at the short wavelength ends of the spectra, with 16 lines 
detected in component A, and 9 lines detected in component B. There are 5 
coincident absorbers, where the wavelengths match within $2.5\sigma$ of the 
measurement errors.  An additional 3 absorbers in B are resolved into 2
components each in A, where the average wavelength in A matches within
the measurement errors to B.  The non-detection of an absorber in A to 
pair to the remaining unmatched line in B is most probably
due to the uncertain continuum fit on the wing of Ly$\alpha$ emission in A.
The strengths of these coincident Ly$\alpha$ absorbers are consistent 
with being equal, given the different signal-to-noise ratios and equivalent 
width limits of the two spectra except for the two matches on the wing of
Ly$\alpha$ emission. This result is expected, since low column
density hydrogen at this redshift has a coherence length much larger than
the transverse separation of these sightlines (Dinshaw et al. 1994; Fang
et al. 1996).

\section{Interpretation of LBQS~0015+0239}

As discussed in the introduction, it can be difficult to directly confirm 
an individual quasar pair as either a lens system or a binary. In this
case, the systemic velocity difference of $661\pm173$ kms$^{-1}$ between
the two components differs from the lensing expectation at the $3.8\sigma$
level. Additional evidence for a binary can come when one image of a 
quasar pair is a stronger radio source than the other (Kochanek et al.
1999). However, LBQS~0015+0239 is not in the area of sky covered by FIRST, 
and NVSS only provides an upper limit to the combined flux density for
both quasars of 2.5 mJy at 1.4 GHz (Condon et al. 1998). The second way 
to search for achromaticity uses the optical spectra. The upper panel of
Figure~2 presents the spectra of the two components with an arbitrary scaling
and offset to illustrate the spectral similarity. The lower panel of Figure~2
shows the ratio of component A to component B where an interpolated continuum 
was used to replace the narrow associated absorption lines in each spectrum.
The flux ratio is roughly a factor of ten with no evidence for any strong
change with wavelength. We fit a straight line to the flux ratio, after
truncating the data below 4400\AA, since the Ly$\alpha$ forest is heavily
absorbed and the flux ratio there will depend on resolution and the
signal-to-noise ratio. The slope is zero to within the error of the fit. 
There are two systematic and significant departures from a flux ratio of 
ten. The first occurs just redward of Ly$\alpha$, reflecting the much 
stronger \ion{N}{5} emission line of component A. The other is centered on 
\ion{C}{4} emission, reflecting different line profiles for the two components. 
The \ion{N}{5} difference between A and B supports the interpretation of the 
pair as a binary, since this difference is much larger than the temporal 
variations seen in individual quasars on timescales appropriate to a time 
delay under the lens hypothesis (Small et al. 1997).

Another approach is to look for a lensing galaxy. No deep image has 
been taken of LBQS~0015+0239 under good enough seeing conditions to put a 
constraint on a lens. However, the Keck spectra are well-resolved spatially, 
so they offer the opportunity of a measurement. The spectra were linearized 
on the final coadded image and then collapsed in wavelength to produce a 1D 
spatial crosscut along a line joining components A and B. Figure~3 shows 
the resulting spatial profile. A Gaussian fit to the two component centers 
in the crosscut yields an image separation of 10.2 pixels or $2.20\pm0.03$ 
arcsec. We choose a wavelength range of 4600-6400\AA\ for the crosscut, 
because it approximates the $V$ passband and it avoids the edges of the 
CCD where the spectra flare out slightly due to defocussing. The bright 
component A has a magnitude of $m_B = 18.7$ on the discovery plate material 
(Hewett, Foltz, \& Chaffee 1995). Direct photometry yields $V = 19.4$ for
component A and $V = 21.9$ for B (Hewett \& Foltz 2002, in preparation). 
Under the lensing hypothesis, the most probable lens redshift is $z = 
0.44$ at a position between A and B but 90\% of the way towards the faint 
component (C. Y. Peng 2002, private communication). We assume an early 
type galaxy with a singular isothermal sphere halo, since known lenses 
generally fit this description (Keeton, Kochanek, \& Falco 1998). At $z 
= 0.44$, the velocity dispersion would be 247 kms$^{-1}$. Scaling from a
local Faber-Jackson relationship predicts a lens luminosity of $1.6L_\ast$, 
which is $V = 20.7$ after (passively) de-evolving the light to $z = 0.44$.

A simple lens model therefore sets the expectation that the lens should be 
located between components A and B, and it should be much brighter than the
faint component. Even without deep imaging, our spectroscopy is sufficient
to rule this out. A Gaussian is actually a poor fit to the wings of the
image profile in Figure~3. Instead, we omit the peak of the profile and 
fit a Gaussian to the lower slope and wing of component A in the direction
away from component B, since a lens will not contribute significant light 
at that position. Assuming symmetry, we fold this profile about the peak
of A and use that shape for the wing of A in the direction of B. Component
B is modelled by the same shape scaled by the ratio of peak fluxes. This 
procedure accounts for essentially all of the flux between the peaks of 
the two components, resulting in a strong limit on residual lens light. We
can use the results to calculate the cross-contamination of the spectra
presented in Figure~1: 6\% of B is light from A and 0.03\% of A is light 
from B. The upper limit on the amount of residual extended light at any 
position between A and B corresponds to $V > 22.8$. 

However, in a realistic lensing scenario, the lens would be 10 times closer 
to B than to A. The amount of residual extended light on or near the fainter 
component B gives a much more stringent upper bound of $V > 24.4$. This 
constraint is only weakened in the unlikely scenario that the lens is 
compact enough to be unresolved and is positioned exactly in front of 
component B. However, the flux ratio in Figure~1 shows no achromatic 
signature from the admixture of red light from an early type galaxy, 
and in fact, the 4000\AA\ break would be seen at any lens redshift in 
the range $0.08 < z < 0.62$. Therefore, any putative lens galaxy might 
be at least 30 times underluminous for its mass compared to galaxies in 
confirmed lens systems. Relative to the stellar mass-to-light ratio of an
early type galaxy, this corresponds to a limits of $(M/L)_V \gtrsim 200$.
However, since the probability distribution for the ``most probable'' lens
redshift is fairly broad, these limits could be a factor of 3-5 weaker.
Since there is no other compelling evidence for massive dark galaxies, 
we prefer to interpret LBQS~0015+0239 as a binary quasar. 

\section{Incidence of Binaries and Lenses in the LBQS}

The LBQS contains 1067 quasars, selected according to their spectral 
energy distributions on objective prism plates from the UK Schmidt
Telescope (Hewett, Foltz \& Chaffee 1995). The survey covers an apparent 
magnitude range of $16.0 \leq m_B \leq 18.85$, a redshift range of $0.2
\leq z \leq 3.4$, and is known to be no more than about 10\% incomplete 
fainter than $m_B = 16.5$ based on a comparison with the FIRST radio 
survey (Hewett, Foltz, \& Chaffee 2001). The LBQS has been subject to 
a specific search for companions within 10 arcsec of each quasar, 
down to level of $m_B = 21.5$ (Hewett et al. 1998). The ongoing 2dF 
and SDSS quasar surveys will eventually produce substantial new samples 
of close pairs. However, in each case, extensive follow-up observations 
are required. Both surveys use fibers which preclude the spectroscopic 
observation of pairs with separations under 20 arcsec. In addition, 
pairs less than 6 arcsec (for the 2dF) and 1.5 arcsec (for SDSS) 
have images that are merged on the original survey material. At present, 
the LBQS is the largest sample that has been closely inspected for pairs.

There are five quasar pairs with separations under 10 arcsec in the 
LBQS. One pair is a confirmed gravitational lens system, and the other 
four are probable binaries. Hewett et al. (1998) described the follow-up 
spectroscopy of plausible candidates for quasar companions. Quasar pairs 
with widely discrepant redshifts are expected to occur by chance. Given 
the surface density of quasars down to $m_B = 21.5$, Hewett et al. (1998) 
predicted 1.6 quasar pairs with unrelated redshifts in the LBQS, which is
consistent with the one known (Hewett et al. 1994). LBQS~1009--0252 is 
confirmed as a lens system at $z = 2.74$, primarily due to the detection 
of a lens galaxy at an appropriate location between the two components 
(Claeskens et al. 2001). In addition, there is metal line absorption in 
the two spectra at $z = 0.87$, which is close to the lens redshift as 
predicted from a model and as estimated from the photometric properties. 
There is also a third image which is a quasar at $z = 1.62$, so this 
system qualifies as a second quasar pair with discrepant redshifts.
The largest surveys for gravitational lenses are radio surveys carried
out at Jodrell Bank and the VLA (Patnaik et al. 1992; Myers et al. 1995; 
Phillips et al. 2001). With 18 lenses from 11,670 radio sources, roughly
half of which are doubles, the presence of only one lens in the LBQS is 
consistent with expectations.

LBQS~1429--0053 is a pair at $z = 2.08$ with an angular separation of 
$\Delta\theta = 5.1$ arcsec (Hewett et al. 1989). Under the lensing
hypothesis, this separation is most likely to be caused by a single 
massive galaxy with an assist from a surrounding cluster. However, 
Hewett et al. find no luminous object at the expected lens position,
with a lower limit of $2000 < (M/L) < 4000$, depending on cosmology.
In addition, the spectra show a difference of $\sim$50\% in the relative 
strength of the Ly$\alpha$-\ion{N}{5} complex, therefore LBQS~1429--008 
is a probably binary. LBQS~2153--2056 is a pair at $z = 1.85$ with an 
angular separation of $\Delta\theta = 7.8$ arcsec (Hewett et al. 1998). 
Available imaging is not deep enough to place a sensitive limit on a 
deflector, but the fact that only one member of the pair is a BAL 
quasar argues against lensing unless the BAL clouds are extremely 
small (see Mortlock et al. 1999). LBQS~0103-2753 is by far the smallest
separation binary known (Junkkarinen et al. 2001), at $z = 0.85$ with
a separation of $\Delta\theta = 0.3$ arcsec. The binary interpretation
is secure because the spectra are very different (and one quasar has 
BAL features). Under the lensing hypothesis, the time delay is $\sim$10
days, and spectral variations sufficient to explain the differences
between two components' spectra are never observed in quasars on such 
a short timescale. As we have argued in this paper, LBQS~0015+0239 is 
a strong candidate for a fourth binary quasar in the LBQS.

The statistical properties of the LBQS pairs provide additional evidence
that they are in fact binaries. Three out of four have flux ratios of ten
or greater, versus only 3 out of 24 two-image lenses with data tabulated
by CASTLES. Lack of spectral similarity provides an additional argument.
The primary of the serendipitously discovered close pair LBQS~0103-2753 
was selected as a BAL quasar, but the companion has a very different, 
and unexceptional, spectrum with no strong absorption. Mortlock et al. 
(1999) used principal compnents analysis to show that LBQS~1429--0053 and 
LBQS~2153--2056 are as dissimilar as any quasar pairs randomly selectly 
from the LBQS. This would not be expected if either pair were lensed. 

The pair that is the subject of this paper is unusual in that both quasars 
show associated absorption. This is noteworthy, given that fewer than 10\% 
of LBQS quasars have this attribute. Associated absorbers are usually 
defined as metal systems within $\Delta v = \pm 5000$ kms$^{-1}$ of the 
emission redshift. Such absorption might be intrinsic and related to the 
broad absorption line (BAL) phenomenon, or extrinsic and caused by galaxies
in a cluster containing the quasar (Foltz et al. 1988). Evidence for 
the latter interpretation comes from abundance studies. Nine out of ten 
associated systems have ($N/C$) ratios below solar abundance, while BAL 
regions have ($N/C$) ratios of 9-10 times solar abundance (Turnshek et 
al. 1996; Franceschini \& Gratton 1997). This new pair also supports the 
extrinsic explanation of associated absorption, as the redshift difference 
of $492\pm6$ kms$^{-1}$ is consistent with a group or cluster environment, 
and the fact that the pair is a binary system shows that there are at 
least two massive galaxies in this region of space.

To date, four binary quasars and a single gravitational lens have been 
discovered in the LBQS. The incidence of binary pairs is difficult to
predict a priori, but it is worth asking what is expected for the rate
of lensing in a survey like the LBQS. Maoz et al. (1993) conducted the 
largest optical lens survey that was sensitive to the angular separations 
of essentially all known lens systems. The survey found 5 lenses out of
502 quasars surveyed (Maoz gave a range of 3-6, so this count includes
0957+561 but excludes the unconfirmed candidate UM~1120+0154), for a
lensing rate of 1\%. For the purposes of this paper, we will not carry
out a full analysis of the predicted lensing rate of the LBQS, but we
can easily give an estimate relative to the Maoz HST snapshot survey.

The predicted number of multiply imaged quasars for any sample is found
by calculating the probability of each quasar being lensed and summing
over these probabilities for the entire sample. The probability that 
an individual quasar is lensed is the product of its optical depth to 
lensing and its magnification bias (Turner, Ostriker, \& Gott 1984;
Fukugita \& Turner 1991). Pei (1995) has published maps of magnification
bias in redshift and apparent blue magnitude, assuming a double power
law model for the quasar luminosity function and a lens population that
is composed of compact objects, individual galaxies, and clusters. The
shape of the magnification bias contours is insensitive to quite varied
models of the lensing population. Taking the 50\% and 90\% bounds in
redshift and magnitude for the two samples, it turns out that the mean
magnification bias is low for both the LBQS and the HST snapshot survey,
just under 1\%. The most substantial difference between the two samples
is the mean redshift, about 1.2 for the LBQS and 2.0 for the HST snapshot
survey. For a $\Lambda$-dominated cosmology, the lensing optical depth
scales as $\tau \sim z^3$ (Bahcall et al. 1992). Thus, the lensing rate
should be 4-5 times lower for the LBQS than for the HST snapshot survey,
consistent with the single lens that has so far been identified.

Despite this argument, it is admitted that the predictions of lensing
rates are uncertain, and the purely observational limits on additional
lenses in the LBQS are poor. The companion search of Hewett et al. (1998) 
was only sensitive to separations greater than 3 arcsec (although equal 
brightness doubles as close as 2 arcsec could have been identified). 
The LBQS search was therefore insensitive to about 90\% of the lens 
image separations from a radio survey such as CLASS. The presence of
a few more small separation lenses in the LBQS cannot be ruled out.

\section{The Population of Binary Quasars}

The LBQS contains four binary pairs with projected separations less than 
$100 h_{70}^{-1}$ kpc (plus the previously mentioned lens which is, of 
course, a single quasar). Since the newly discovered LBQS~0015+0239 has 
smaller angular separation than the 3-10 arcsec completeness range of 
the search by Hewett et al. (1998), and since the 0.3 arcsec pair 
LBQS~0103-2753 was discovered serendipitously (Junkkarinen et al. 2001), 
it is worth asking how many more small separation binaries might be 
lurking in the LBQS. Maoz et al. (1993) performed a snapshot survey of 
498 quasars with $z > 1$ and found no binaries with separations below 
3 arcsec. They could have detected binaries as close as 0.3 arcsec, and 
with flux ratios of at least 4:1. Thus, the single close LBQS binary is 
consistent with the null result of Maoz et al., and Poisson statistics
suggest that there are unlikely to be more than two additional very close
binaries in the LBQS. 

Over the order of magnitude larger range of scales
from 10 to 100 arcsec, there are no LBQS pairs with similar redshifts. 
Unfortunately, this null result is not a strong constraint, since the 
deep companion search of Hewett et al. (1998) applied over this angular 
range would have been been sensitive to 30 times as many quasars. As 
noted originally by Djorgovski (1991), the high incidence of close pairs 
is strong circumstantial evidence that quasar activity can be triggered by
direct interactions. Extrapolating the observed quasar correlation function
down to these small physical scales leads to a prediction of $\sim$ 0.01
quasar pairs within $100 h_{70}^{-1}$ kpc, compared to the four observed.
A similar scaling of the correlation function predicts 20 times as many 
pairs between 10 and 100 arcsec as between 3 and 10 arcsec, or $\sim$40 
wide pairs. 

We adopt the pair fraction of the LBQS (4/1055) as the probability that a 
quasar is a member of a binary, $P_{binary} = 4 \times 10^{-3}$. The LBQS
probability only applies out to 10 arcsec, or a transverse separation of
$\sim 100 h_{70}^{-1}$ kpc. If the interaction-induced quasar activity 
occurs in the advanced stages of a merger process, then we anticpate
that the probability summed to much larger separation scales will not be
much larger (Barnes \& Hernquist 1996). If the radio-loud fraction of 
optically selected quasars is $f_{rad}$ and we assume that binarism does 
not affect radio properties, then the relative numbers of $O^2$, $O^2R$, 
and $O^2R^2$ binaries (ie. those where both, one, and neither quasar are 
radio-loud) are in the ratio $(1-f_{rad})^2$ : $2f_{rad}(1-f_{rad})$ : 
$f_{rad}^2$ (Kochanek et al. 1999). Updating the Mortlock et al. list 
(see comments in the Introduction) gives 12 $O^2$ pairs. Taking $f_{rad} 
\simeq 0.1$, as is appropriate for FIRST or NVSS detections above 1-3 mJy, 
the predictions are 0.1 $O^2R^2$ pair, which is consistent with none seen, 
and 2.2 $O^2R$, which is consistent with two seen. 

Recent radio lens surveys are an order of magnitude larger than the LBQS.
Also, the surface density of radio sources is low enough that it is much
more feasible to search for pairs out to larger angular separations. So 
far, only one radio-loud binary ($O^2R^2$) has been found among 15,000 
CLASS sources (Koopmans et al. 2000), and none have been found among 9100 
FIRST sources (Ofek et al. 2002). The only other known $O^2R^2$ binary is 
MG~0023+171 (Hewitt et al. 1987). It is entirely plausible that several 
more $O^2R^2$ binaries remain to be found since the radio morphologies 
of each component may be quite different, thus eluding the selection 
designed to catch lens candidates. Using the LBQS estimate of $P_{binary}$, 
the total number of $O^2R$ binaries present in the CLASS and FIRST 
surveys out to $\Delta\theta = 10$ arcsec should be $\sim$57 and 
$\sim$35, respectively. The number where both components are radio-loud 
is a factor of $f_{rad}$ smaller, or $\sim$3 and $\sim$4, respectively. 
The numbers should also be increased to account for the larger angular
scales of the CLASS (15 arcsec) and FIRST (30 arcsec) surveys, assuming
that the association probability scales according to the quasar-galaxy
correlation function. The smaller size of the FIRST survey is outweighed
by the deeper radio limit and the larger angular search radius. For the
FIRST survey, we predict $\sim$15 $O^2R$ binaries. We predict $\sim$100
and $\sim$400 binaries in the 2dF and SDSS optical quasar surveys, but
note that substantial spectroscopic followup will be needed to identify
the second (fainter) components in each survey.

LBQS~0015+0239 is the third smallest separation binary quasar known. 
Kochanek et al. (1999) have shown that number of binaries is consistent 
with the quasar-galaxy correlation function at small scales. Mortlock 
et al. (1999) deprojected the observed separations of binary pairs 
to yield a broad distribution peaking at $\sim50 h_{70}^{-1}$ kpc, 
a declining tail out to large separations, and a possible deficit at 
small separations. Despite the lack of comprehensive follow-up of wide
pairs from radio surveys, the decline at large separations appears to
be real and is indicative of an interaction trigger for the nuclear
activity. LBQS~0015+0239 and the other two binary quasars discovered
since Mortlock et al. neatly fill in their ``hole,'' so the separation
distribution may be flat below $50 h_{70}^{-1}$ kpc. Given this type of
distribution, the 0.3 arcsec LBQS pair discovered by Junkkarinen et al.
(2001) is far more likely to be a real 2-3 $h_{70}^{-1}$ binary than a
chance projection of a $\sim40 h_{70}^{-1}$ kpc binary. LBQS quasars are 
typically found in massive galaxies (Hooper, Impey \& Foltz 1997), where 
the dynamical decay timescale of the binary orbits is close to a Hubble 
time and much longer than active timescale of quasars accreting at near
the Eddington limit (Binney \& Tremaine 1987). Thus, LBQS~0015+0239 may 
have been triggered by the much faster settling process with the merged 
halo (Barnes \& Hernquist 1996). Deep imaging of this pair, and many of 
the other binaries, should reveal heavily distorted host morphologies.

\section{Conclusions}

We have carried out Keck/LRIS spectroscopy of the 2.2 arcsec separation
quasar pair LBQS~0015+0239. These data, plus a consideration of the other
pairs in the LBQS, leads to the following conclusions:

1. There are three pieces of evidence that support the interpretation
that the pair is a binary system. The velocity difference between the 
components is inconsistent with the lensing expectation but consistent 
with a bound pair of massive galaxies. There are substantial differences 
in \ion{N}{5}, and to a lesser extent \ion{C}{4}, emission line strengths 
between the two components. A spatial cut along the line connecting the pair 
in the equivalent of the $V$ band leads to the conclusion that any lensing 
galaxy must be at least thirty times less luminous than an elliptical galaxy 
that could account for the image separation.

2. Narrow associated absorption due to metal lines is seen in each
component, with a velocity difference of $492\pm6$ kms$^{-1}$. Since the
associated absorption phenomenon is only seen in about 10\% of LBQS
spectra, its presence in these two quasars supports the idea that the
absorption arises in a group or cluster environment that includes the 
binary pair.

3. There are four probable binary pairs in the LBQS, so the incidence
rate is $4 \times 10^{-3}$, several times higher than the lensing rate.
The large radio surveys being used to select gravitational lenses probably
contain 15-20 unrecognized binary pairs, where only one of the members is
radio-loud. LBQS~0015+0239 has a projected transverse separation of only
17.8 $h_{70}^{-1}$ kpc and is the highest redshift binary quasar known.
It may represent the short-lived phase prior to the coalescence of two 
supermassive black holes, in which case deep imaging of the pair should 
reveal a disturbed morphology for the host galaxies.

\acknowledgements

We are grateful to the excellent staff of the W. M. Keck Observatory for 
facilitating these observations. We benefited from useful discussions on 
the debate over lensing versus binarism with Chris Kochanek, Ian Browne,
and Chien Peng. We thank the referee for a number of useful suggestions.
CDI and CBF acknowledge support from the NSF under award AST-9803072.

\clearpage

\references

\reference{bah92} Bahcall, J. N., Maoz, D., Doxsey, R., Schneider, D. P.,
	Bahcall, N. A., Lahav, O., \& Yanny, B. 1992, \apj, 387, 56
\reference{bar96} Barnes, J., \& Hernquist, L. 1996, \apj, 471, 115
\reference{bin87} Binney, J., \& Tremaine, S. 1987, Galactic Dynamics,
	Princeton University Press: Princeton
\reference{bis97} Bischof, O. B., \& Becker, R. H. 1997, \aj, 113, 2000
\reference{bla92} Blandford, R. D., \& Narayan, R. 1992, \araa, 30, 311
\reference{cla01} Claeskens, J.-F., Khmil, S. V., Lee, D. W., Sluse, D.,
	\& Surdej, J. 2001, \aap, 367, 748
\reference{con98} Condon, J. J., Cotton, W. D., Greisen, E. W., Yin, Q. F.,
	Perley, R. A., et al. 1998, \aj, 115, 1693
\reference{cro01} Croom, S. M., Smith, R. J., Boyle, B. J., Shanks, T. 
	\& Loaring, N. S. 2001, \mnras, 322, L29
\reference{din94} Dinshaw, N., Impey, C. D., Foltz, C. B., Weymann, R. J.,
	\& Chaffee, F. C. 1994, \apjl, 437, L87
\reference{djo91} Djorgovski, S. 1991, in ASP Conference Series Vol. 21
	on the Space Distribution of Quasars, ed. D. Crampton, ASP: San
	Francisco, p. 349
\reference{fal99} Falco, E. E., Impey, C. D., Kochanek, C. S., L\'ehar,
	J., McLeod, B. A. et al. 1999, \apj, 523, 617
\reference{fan96} Fang, Y., Duncan, R. C., Crotts, A.P S., \& Bechtold, J.
	1996, \apj, 462, 77
\reference{fol88} Foltz, C. B., Chaffee, F. H., Weymann, R. J., \& 
	Anderson, S. J. 1988, in QSO Absoprtion Lines: Probing the Universe,
	eds. C. Blades, D. turnshek, \& C. Norman, Cambridge: Cambridge
	University Press, p. 66
\reference{fra97} Franceschini, A., \& Gratton, R. 1997, \mnras, 286, 235
\reference{fra92} Francis, P. J., Hewett, P. C., Foltz, C. B., \& Chaffee,
	F. C. 1992, \apj, 398, 476
\reference{gre02} Green, P. J., Kochanek, C. S., Siemiginowska, A., Kim,
	D.-W., Markevitch, M. et al. 2002, \apj, in preparation
\reference{haw97} Hawkins, M. R. S. 1997, \aap, 328, L25
\reference{hew89} Hewett, P. C., Webster, R. L., Harding, M. E., 
	Jedrzejewski, R. I., Foltz, C. B., et al. 1989, \apj, 346, L61
\reference{hew94} Hewett, P. C., Irwin, M. J., Foltz, C. B., Harding, M. 
	E., Corrigan, R. T., Webster, R. L., \& Dinshaw, N. 1994, \aj, 
	108, 1534
\reference{hew95} Hewett, P. C., Foltz, C. B., \& Chaffee, F. H. 1995,
	\aj, 122, 518
\reference{hew98} Hewett, P. C., Foltz, C. B., Harding, M. E., \& Lewis,
	G. F. 1998, \aj, 115, 383
\reference{hew01} Hewett, P. C., Foltz, C. B., \& Chaffee, F. H. 2001,
	\apj, 122, 518
\reference{hew87} Hewitt, J. N., Turner, E. L., Lawrence, C. R., Schneider,
	D. P., Gunn, J. E., et al. 1987, \apj, 321, 706
\reference{hoo96} Hooper, E. J., Impey, C. D., Foltz, C. B., \& Hewett, 
	P. C. 1996, \apj, 473, 746
\reference{hoo97} Hooper, E. H., Impey, C. D., \& Foltz, C. B. 1997, \apjl,
	480, L95
\reference{fuk91} Fukugita, M., \& Turner, E. L. 1991, \mnras, 253, 99
\reference{jun01} Junkkarinen, V., Shields, G. A., Beaver, E. A., Burbidge,
	E. M., Cohen, R. D. et al. 2001, \apjl, 549, L155
\reference{kee98} Keeton, C. R., Kochanek, C. S., \& Falco, E. E. 1998,
	\apj, 509, 561
\reference{koc95} Kochanek, C. S. 1995, \apj, 453, 545
\reference{koc99} Kochanek, C. S., Falco, E. E., \& Mu\~noz, J. A. 1999,
	\apj, 510, 590
\reference{koo00} Koopmans, L. V. E., de Bruyn, A. G., Fassnacht, C. D.,
	Marlow, D. R., Rusin, D. et al. 2000, \aap, 361, 822
\reference{law84} Lawrence, C. R., Schneider, D. P., Schmidt, M., Bennett,
	C. L., Hewitt, J. N. et al. 1984, Science, 223, 46
\reference{leh00} L\'ehar, J., Falco, E. E., Kochanek, C. S., McLeod, B. 
	A., M\~unoz, J. A. et al. 2000, \apj, 536, 584
\reference{lew98} Lewis, G. F., Irwin, M. J., Hewett, P. C., \& Foltz, C. B.,
	1998, \mnras, 295, 573
\reference{mao93} Maoz, D., Bahcall, J. N., Schneider, D. P., Bahcall, N. 
	A., Djorgovski, S. 1993, \apj, 409, 28
\reference{mor00} Morgan, N. D., Burley, G., Costa, E., Maza, J., Persson,
	S. E. et al. 2000, \aj, 119, 1083
\reference{mor99} Mortlock, D. J., Webster, R. L., \& Francis, P. J. 1999,
	\mnras, 309, 836
\reference{mye95} Myers, S. T., Fassnacht, C. D., Djorgovski, S. G., 
	Blandford, R. D., Matthews, K. et al. 1995, \apjl, 447, L5
\reference{ofe01} Ofek, E. O., Maoz, D., Prada, F., Kolatt, T., and Rix,
	H.-W. 2002, \mnras, in press
\reference{pat92} Patnaik, A. R., Browne, I. W. A., Wilkinson, P. N.,
	\& Wrobel, J. M. 1992, \mnras, 254, 655
\reference{pei95} Pei, Y. C. 1995, \apj, 440, 485
\reference{pen99} Peng, C. Y., Impey, C. D., Falco, E. E., Kochanek, C. S.,
	L\'ehar, J. et al. 1999, \apj, 524, 572
\reference{pet98} Petry, C. E., Impey, C. D., \& Foltz, C. B. 1998, \apj,
	494, 60
\reference{pet02} Petry, C. E., Impey, C. D., Katz, N., Weinberg, D. H., 
	\& Hernquist, L. E. 2002, \apj, in press
\reference{phi01} Phillips, P. M., Browne, I. W. A., Jackson, N. J.,
	Wilkinson, P. N., Mao, S. et al. 2001, \mnras, 328, 1001
\reference{ric01} Richards, G. T., Fan, X., Schneider, D. P., Vanden Berk,
	D. E., Strauss, M. A., et al. 2001, \aj, 121, 2308
\reference{rod94} Rodgers, J. M., \& McCarthy, J. K. 1994, SPIE, 2198, 1096
\reference{sch92} Schneider, P., Ehlers, J., \& Falco, E. E. 1992, 
	Gravitational Lenses, Berlin: Springer-Verlag
\reference{sch93} Schneider, D. P., Hartig, G. F., Jannuzi, J. P., Kirhakos,
	S., Saxe, D. H., et al. 1993, \apjs, 87, 45
\reference{sma97} Small, T. A., Sargent, W. L. W., \& Steidel, C. C. 1997,
	\aj, 114, 2254
\reference{ste91} Steidel, C. C. \& Sargent, W. L. W. 1991, \aj, 102, 1610
\reference{sur02} Surdej, J., \& Claeskens, J.-F. 2002, \araa, in press
\reference{tur84} Turner, E. L., Ostriker, J. P., \& Gott, J. R. 1984,
	\apj, 284, 1
\reference{tur96} Turnshek, D. A., Kopko, M., Minier, E., Noll, D, Espey,
	D. R., \& Weymann, R. J. 1996, \apj, 463, 110
\reference{van01} Vanden Berk, D. E., Richards, G. T., Bauer, A., Strauss,
	M. A., Schneider, D. P. et al. 2001, \aj, 122, 549 
\reference{wal79} Walsh, D., Carswell, R. F., \& Weymann, R. J. 1979,
	Nature, 279, 381
\reference{wis93} Wisotzki, L., Kohler, T., Kayser, R., \& Reimers, D.
	1993, \aap, 278, L15

% Figures
\begin{figure}
\epsscale{0.8}
\plotone{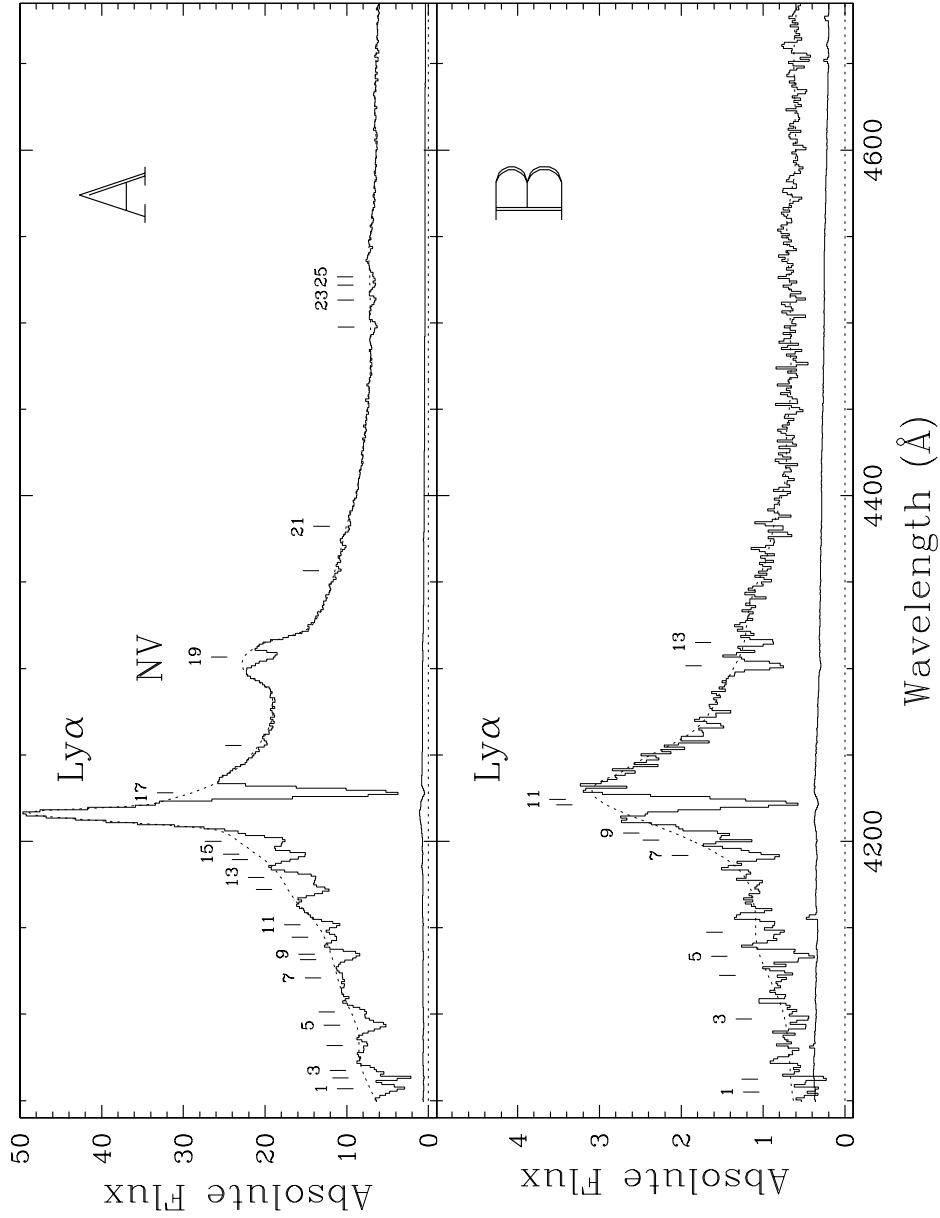}
\caption{Keck/LRIS spectra of the quasar pair LBQS~0015+0239, with 
4.2\AA\ resolution.  Cross correlation with SDSS quasar composite gives
redshifts of $z=2.4747$ for component A and $z=2.4831$ for component B.
The dotted line is the fitted continuum, prominent emission features are
labelled, and all significant absorption lines are marked with every other 
one numbered. The lower curve shows the $3\sigma$ flux error.
The flux is in units of 10$^{-17}$ ergs s$^{-1}$ cm$^{-2}$ \AA$^{-1}$.
}
\end{figure}

\begin{figure}
\epsscale{0.8}
\plotone{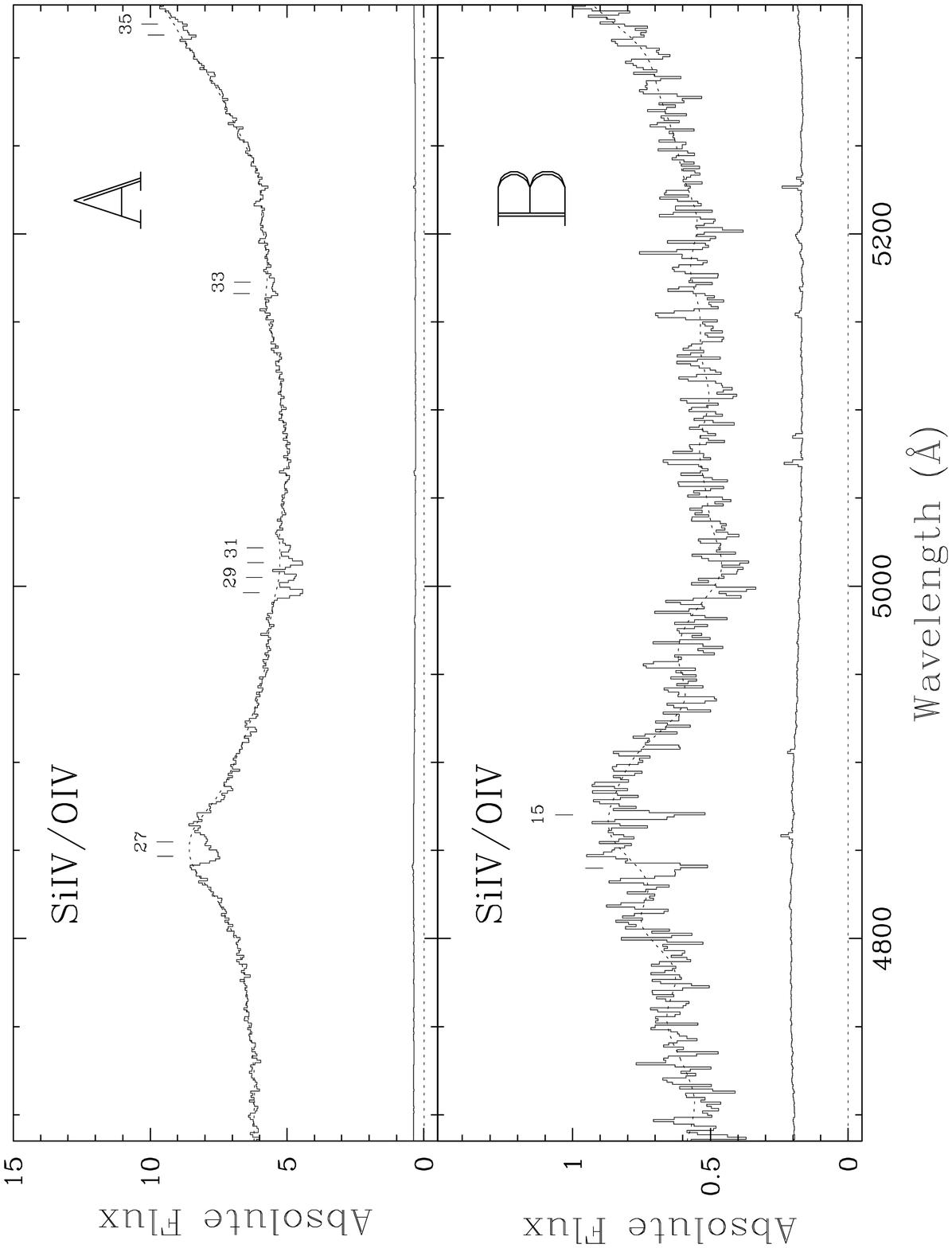}
\end{figure}

\begin{figure}
\epsscale{0.8}
\plotone{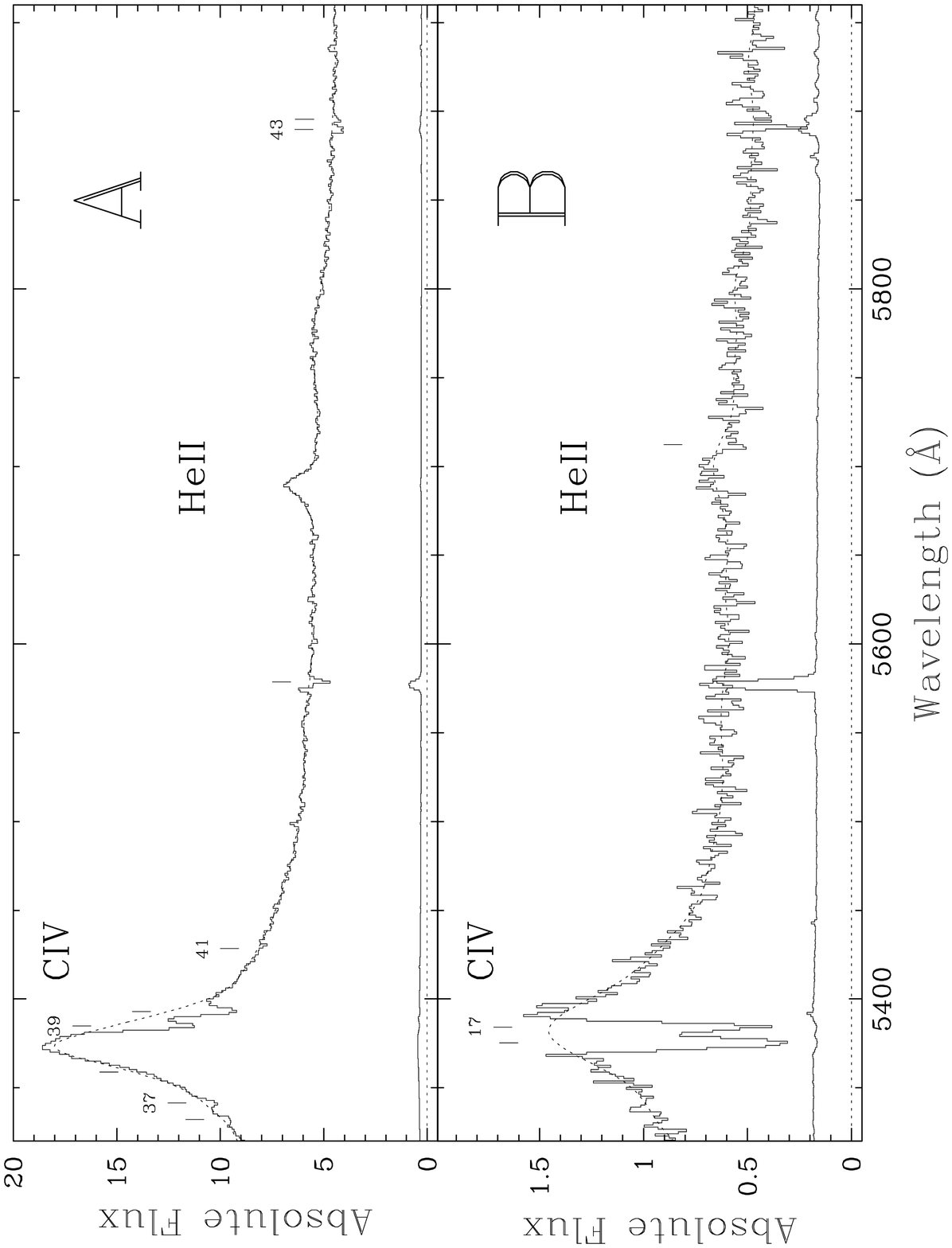}
\end{figure}

\begin{figure}
\epsscale{0.8}
\plotone{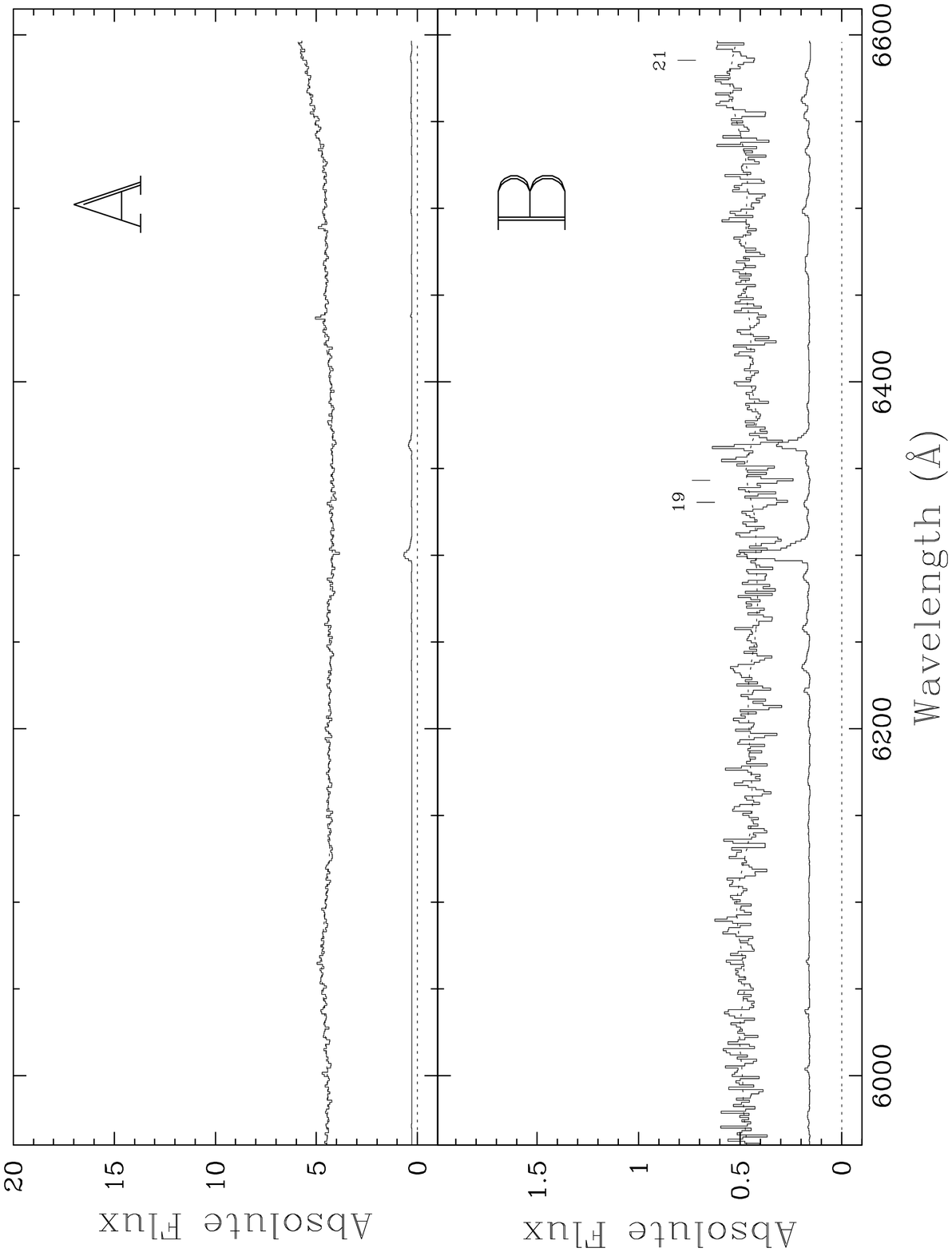}
\end{figure}

\begin{figure}
\epsscale{0.8}
\plotone{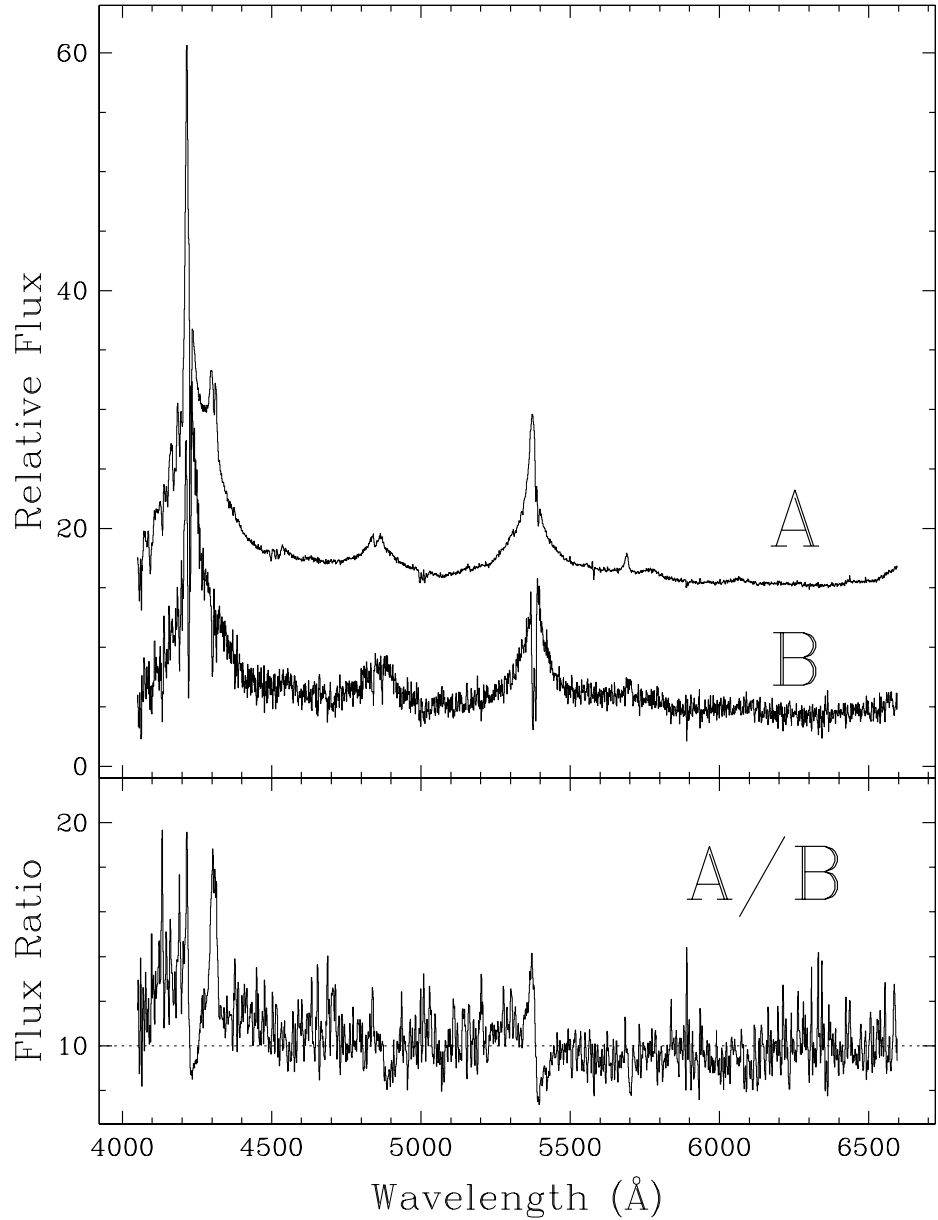}
\caption{
The upper panel shows the spectra of LBQS~0015+0239 components A and B scaled
and offset by an arbitrary amount. The lower panel shows
the ratio of component A to component B after interpolation
over the narrow associated absorption features.  The dotted line is plotted at
the approximate mean flux ratio.
}
\end{figure}

\begin{figure}
\epsscale{0.8}
\plotone{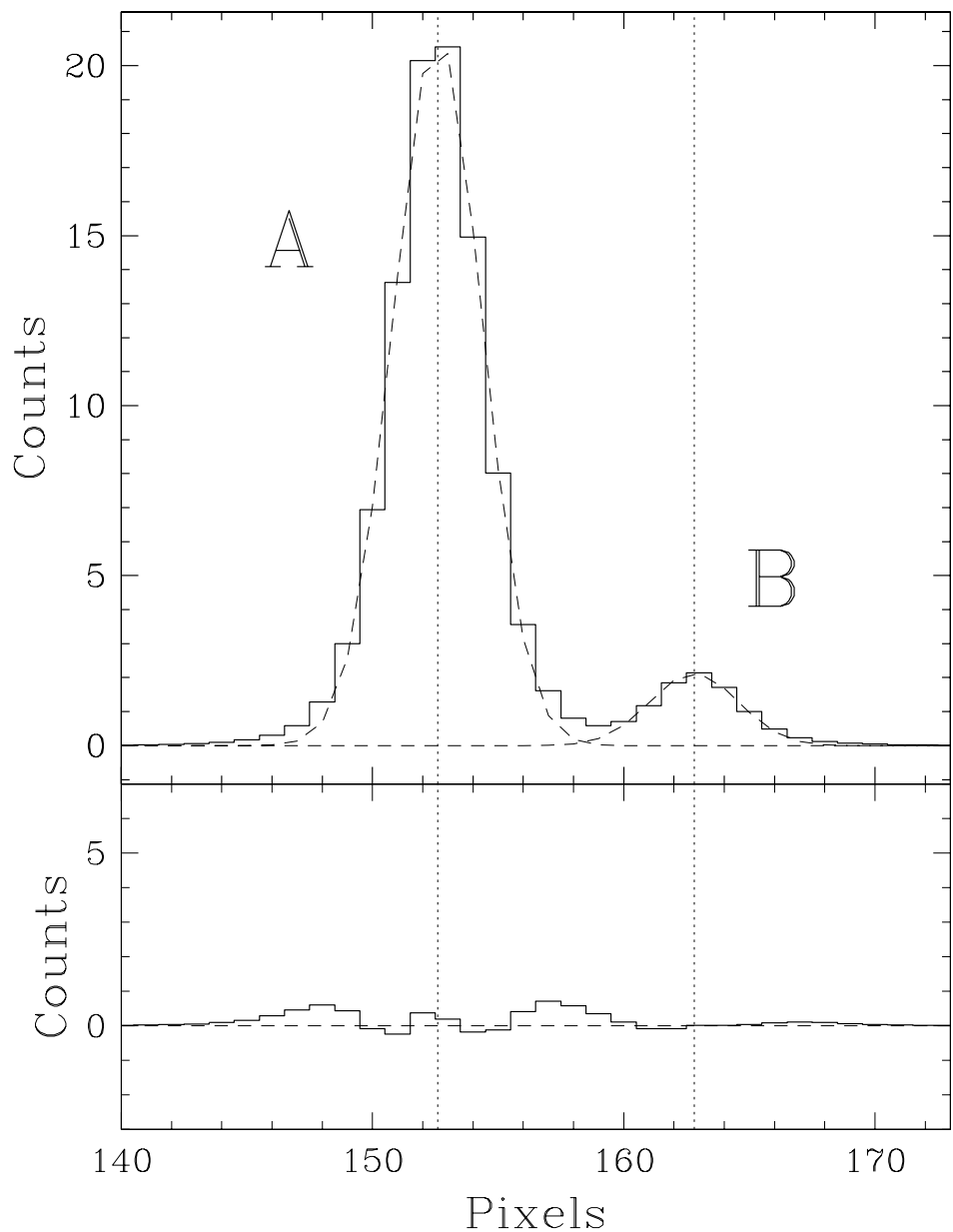}
\caption{
The upper panel shows the spatial cross-cut of LBQS~0015+0239 over the spectral
range 4600-6400\AA.  The fitted profile (dashed line) is scaled from the 
left-hand side of component A, where any potential light from a lensing galaxy 
will be negligible.  The lower panel shows the residual counts after subtraction
of the two fitted components.  In both panels the counts are in units of 
$10^{4}$, and the vertical dotted lines show the centers of the two components.
}
\end{figure}

% Tables
\clearpage
%\renewcommand{\baselinestretch}{1.0}

% Table 1
\begin{deluxetable}{rcccrrr} 
   
\tablenum{1} 
\tablewidth{0pt} 
\tablecaption{LBQS~0015+0239 A}
\tablehead{ 
 \colhead{Line}  & 
 \colhead{Central $\lambda$\tablenotemark{a}}  & 
 \colhead{Equivalent Width}  & 
 \colhead{FWHM\tablenotemark{b}} & 
 \colhead{$\chi^{2}_{\nu}$} & 
 \colhead{SL$_{fit}$\tablenotemark{c}} & 
 \colhead{SL$_{det}$\tablenotemark{d}} \\ 
 \colhead{}  & 
 \colhead{(\AA)}  & 
 \colhead{(\AA)}  & 
 \colhead{(\AA)}  & 
 \colhead{}  & 
 \colhead{}  & 
 \colhead{}  
}   
\startdata 
   1 & 4056.80 $\pm$ 0.08 & 2.680 $\pm$ 0.110 & 4.37 $\pm$ 0.21 & 17.25 & 24.70 & 31.33  \\ 
   2 & 4063.23 $\pm$ 0.17 & 2.600 $\pm$ 0.180 & 4.20 $\pm$ 0.00 & 17.25 & 14.70 & 32.92  \\ 
   3 & 4067.52 $\pm$ 0.68 & 0.720 $\pm$ 0.230 & 4.69 $\pm$ 1.38 & 17.25 &  3.10 &  9.45  \\ 
   4 & 4081.93 $\pm$ 0.36 & 0.570 $\pm$ 0.090 & 4.56 $\pm$ 0.85 &  2.25 &  6.20 &  7.90  \\ 
   5 & 4093.57 $\pm$ 0.14 & 2.610 $\pm$ 0.130 & 6.12 $\pm$ 0.37 &  0.72 & 19.80 & 37.10  \\ 
   6 & 4101.25 $\pm$ 0.26 & 1.050 $\pm$ 0.120 & 5.01 $\pm$ 0.62 &  0.72 &  8.80 & 15.97  \\ 
   7 & 4121.04 $\pm$ 0.71 & 0.460 $\pm$ 0.100 & 6.87 $\pm$ 1.84 &  1.45 &  4.50 &  8.05  \\ 
   8 & 4131.63 $\pm$ 2.26 & 0.640 $\pm$ 0.590 & 5.69 $\pm$ 3.30 &  1.45 &  1.10 & 11.82  \\ 
   9 & 4134.77 $\pm$ 0.24 & 1.170 $\pm$ 0.540 & 4.20 $\pm$ 0.00 &  1.45 &  2.20 & 21.88  \\ 
  10 & 4144.57 $\pm$ 0.22 & 0.590 $\pm$ 0.060 & 4.38 $\pm$ 0.55 &  0.35 &  9.40 & 11.52  \\ 
  11 & 4151.71 $\pm$ 0.13 & 0.920 $\pm$ 0.040 & 4.20 $\pm$ 0.00 &  0.35 & 20.40 & 19.27  \\ 
  12 & 4172.03 $\pm$ 0.18 & 1.950 $\pm$ 0.120 & 6.57 $\pm$ 0.41 &  1.70 & 16.70 & 48.23  \\ 
  13 & 4179.14 $\pm$ 0.18 & 1.340 $\pm$ 0.110 & 5.32 $\pm$ 0.38 &  1.70 & 12.20 & 35.02  \\ 
  14 & 4189.52 $\pm$ 0.36 & 0.530 $\pm$ 0.090 & 4.20 $\pm$ 0.00 &  1.55 &  6.10 & 13.12  \\ 
  15 & 4192.63 $\pm$ 0.19 & 1.120 $\pm$ 0.080 & 4.20 $\pm$ 0.00 &  1.55 & 13.90 & 29.14  \\ 
  16 & 4200.14 $\pm$ 0.11 & 1.920 $\pm$ 0.060 & 7.13 $\pm$ 0.28 &  1.55 & 31.20 & 55.27  \\ 
  17\tablenotemark{e} & 4228.10 $\pm$ 0.02 & 5.500 $\pm$ 0.030 & 5.76 $\pm$ 0.04 & 26.80 & 171.10 & 199.65  \\ 
  18\tablenotemark{e} & 4255.52 $\pm$ 0.59 & 0.240 $\pm$ 0.050 & 5.86 $\pm$ 1.42 &  0.72 &  4.90 &  7.05  \\ 
  19 & 4306.59 $\pm$ 0.12 & 1.230 $\pm$ 0.050 & 6.72 $\pm$ 0.29 &  2.78 & 26.90 & 41.53  \\ 
  20 & 4356.66 $\pm$ 0.59 & 0.200 $\pm$ 0.050 & 4.20 $\pm$ 0.00 &  1.26 &  4.30 &  4.30  \\ 
  21 & 4382.18 $\pm$ 0.73 & 0.290 $\pm$ 0.070 & 5.98 $\pm$ 1.77 &  0.53 &  4.00 &  5.77  \\ 
  22 & 4497.70 $\pm$ 0.33 & 0.650 $\pm$ 0.080 & 5.47 $\pm$ 0.79 &  0.77 &  8.10 & 11.13  \\ 
  23 & 4513.08 $\pm$ 0.34 & 0.420 $\pm$ 0.060 & 4.20 $\pm$ 0.00 &  0.80 &  7.50 &  7.31  \\ 
  24 & 4521.86 $\pm$ 0.72 & 0.360 $\pm$ 0.130 & 4.20 $\pm$ 0.00 &  0.48 &  2.70 &  6.29  \\ 
  25 & 4526.61 $\pm$ 1.00 & 0.410 $\pm$ 0.160 & 5.38 $\pm$ 2.17 &  0.48 &  2.50 &  7.20  \\ 
  26\tablenotemark{e} & 4846.76 $\pm$ 0.67 & 0.910 $\pm$ 0.250 & 6.89 $\pm$ 1.17 &  1.06 &  3.60 & 21.07  \\ 
  27 & 4854.93 $\pm$ 1.54 & 0.730 $\pm$ 0.260 & 9.15 $\pm$ 2.78 &  1.06 &  2.80 & 16.93  \\ 
  28 & 4996.39 $\pm$ 0.16 & 0.960 $\pm$ 0.070 & 4.51 $\pm$ 0.38 &  1.39 & 14.00 & 17.28  \\ 
  29 & 5004.91 $\pm$ 0.24 & 0.570 $\pm$ 0.050 & 4.20 $\pm$ 0.00 &  1.39 & 10.60 & 10.15  \\ 
  30 & 5013.37 $\pm$ 0.19 & 0.720 $\pm$ 0.050 & 4.20 $\pm$ 0.00 &  1.07 & 13.20 & 12.76  \\ 
  31 & 5021.86 $\pm$ 0.42 & 0.380 $\pm$ 0.070 & 4.54 $\pm$ 0.99 &  1.07 &  5.30 &  6.79  \\ 
  32 & 5166.22 $\pm$ 0.53 & 0.300 $\pm$ 0.060 & 4.20 $\pm$ 0.00 &  0.41 &  5.30 &  5.88  \\ 
  33 & 5172.70 $\pm$ 0.74 & 0.270 $\pm$ 0.080 & 5.33 $\pm$ 1.91 &  0.41 &  3.40 &  5.27  \\ 
  34 & 5312.80 $\pm$ 0.53 & 0.230 $\pm$ 0.050 & 4.20 $\pm$ 0.00 &  0.84 &  5.00 &  6.10  \\ 
  35 & 5319.14 $\pm$ 0.68 & 0.260 $\pm$ 0.070 & 5.75 $\pm$ 1.75 &  0.84 &  4.00 &  7.08  \\ 
  36 & 5332.23 $\pm$ 0.53 & 0.170 $\pm$ 0.040 & 4.20 $\pm$ 1.27 &  1.16 &  3.90 &  4.82  \\ 
  37 & 5341.45 $\pm$ 0.35 & 0.240 $\pm$ 0.030 & 4.20 $\pm$ 0.00 &  0.36 &  7.30 &  7.12  \\ 
  38 & 5358.85 $\pm$ 0.77 & 0.330 $\pm$ 0.050 & 9.78 $\pm$ 1.95 &  0.44 &  6.10 & 11.56  \\ 
  39\tablenotemark{e} & 5384.81 $\pm$ 0.06 & 1.280 $\pm$ 0.040 & 4.39 $\pm$ 0.14 &  2.34 & 35.10 & 50.63  \\ 
  40\tablenotemark{e} & 5392.90 $\pm$ 0.08 & 1.670 $\pm$ 0.050 & 5.98 $\pm$ 0.21 &  2.34 & 35.10 & 53.67  \\ 
  41 & 5428.49 $\pm$ 3.05 & 0.240 $\pm$ 0.140 & 11.07 $\pm$ 7.76 &  1.02 &  1.70 &  6.16  \\ 
  42 & 5578.61 $\pm$ 0.49 & 0.510 $\pm$ 0.120 & 4.20 $\pm$ 0.00 &  2.16 &  4.20 &  5.02  \\ 
  43 & 5889.74 $\pm$ 0.36 & 0.520 $\pm$ 0.070 & 4.20 $\pm$ 0.00 &  1.03 &  7.00 &  8.68  \\ 
  44 & 5895.53 $\pm$ 0.60 & 0.290 $\pm$ 0.070 & 4.20 $\pm$ 0.00 &  1.03 &  4.30 &  4.85  \\ 
\tablenotetext{a}{Wavelengths are vacuum heliocentric.} 
\tablenotetext{b}{Lines with a value of $4.20 \pm 0.00$ were fit with the FWHM set to the minimum allowed value.} 
\tablenotetext{c}{Significance of the line defined as $EW/\sigma_{fit}$, where $\sigma_{fit}$ is the error in the measured equivalent width.} 
\tablenotetext{d}{Significance of the line defined as $EW/\sigma_{det}$, where $\sigma_{det}$ is the $1\sigma$ limiting equivalent width.} 
\tablenotetext{e}{This line has been identified with an associated absorption
system and is included in Table 3.}
\enddata 
\end{deluxetable} 
  
% Table 2 
\begin{deluxetable}{rcccrrr} 
   
\tablenum{2} 
\tablewidth{0pt} 
\tablecaption{LBQS~0015+0239 B} 
\tablehead{ 
 \colhead{Line}  & 
 \colhead{Central $\lambda$\tablenotemark{a}}  & 
 \colhead{Equivalent Width}  & 
 \colhead{FWHM\tablenotemark{b}} & 
 \colhead{$\chi^{2}_{\nu}$} & 
 \colhead{SL$_{fit}$\tablenotemark{c}} & 
 \colhead{SL$_{det}$\tablenotemark{d}} \\
 \colhead{}  & 
 \colhead{(\AA)}  & 
 \colhead{(\AA)}  & 
 \colhead{(\AA)}  & 
 \colhead{}  & 
 \colhead{}  & 
 \colhead{}  
}   
\startdata 
   1 & 4055.07 $\pm$ 0.79 & 2.730 $\pm$ 0.820 & 5.56 $\pm$ 2.02 &  1.07 &  3.30 &  5.11  \\ 
   2 & 4062.30 $\pm$ 0.65 & 2.390 $\pm$ 0.560 & 4.20 $\pm$ 0.00 &  1.07 &  4.20 &  4.42  \\ 
   3 & 4097.22 $\pm$ 1.21 & 2.890 $\pm$ 0.810 & 8.87 $\pm$ 2.92 &  1.16 &  3.60 &  6.40  \\ 
   4 & 4122.33 $\pm$ 0.97 & 1.590 $\pm$ 0.530 & 6.01 $\pm$ 2.37 &  0.35 &  3.00 &  4.46  \\ 
   5 & 4133.43 $\pm$ 0.37 & 4.310 $\pm$ 0.500 & 6.68 $\pm$ 0.94 &  1.10 &  8.60 & 13.52  \\ 
   6 & 4147.43 $\pm$ 0.97 & 2.700 $\pm$ 0.580 & 9.62 $\pm$ 2.55 &  0.92 &  4.70 &  9.13  \\ 
   7 & 4191.85 $\pm$ 0.32 & 2.740 $\pm$ 0.310 & 5.74 $\pm$ 0.77 &  1.49 &  8.80 & 12.59  \\ 
   8 & 4200.63 $\pm$ 0.55 & 1.170 $\pm$ 0.210 & 4.20 $\pm$ 0.00 &  1.49 &  5.50 &  6.37  \\ 
   9 & 4204.79 $\pm$ 0.50 & 1.130 $\pm$ 0.210 & 4.20 $\pm$ 0.00 &  1.49 &  5.40 &  6.92  \\ 
  10\tablenotemark{e} & 4221.22 $\pm$ 0.58 & 2.910 $\pm$ 0.350 & 4.20 $\pm$ 0.00 &  0.88 &  8.40 & 24.42  \\ 
  11 & 4224.33 $\pm$ 0.41 & 1.930 $\pm$ 0.440 & 4.20 $\pm$ 0.00 &  0.88 &  4.40 & 16.68  \\ 
  12\tablenotemark{e} & 4301.56 $\pm$ 0.26 & 2.150 $\pm$ 0.260 & 4.34 $\pm$ 0.61 &  0.64 &  8.20 &  9.91  \\ 
  13\tablenotemark{e} & 4315.10 $\pm$ 0.51 & 1.180 $\pm$ 0.300 & 4.20 $\pm$ 1.29 &  1.56 &  3.90 &  4.99  \\ 
  14\tablenotemark{e} & 4839.88 $\pm$ 0.62 & 1.540 $\pm$ 0.370 & 5.44 $\pm$ 1.56 &  1.21 &  4.20 &  6.03  \\ 
  15\tablenotemark{e} & 4870.19 $\pm$ 0.35 & 1.560 $\pm$ 0.210 & 4.20 $\pm$ 0.00 &  0.77 &  7.40 &  7.17  \\ 
  16\tablenotemark{e} & 5375.40 $\pm$ 0.13 & 6.370 $\pm$ 0.250 & 7.30 $\pm$ 0.35 &  2.42 & 25.50 & 50.60  \\ 
  17\tablenotemark{e} & 5384.23 $\pm$ 0.13 & 4.080 $\pm$ 0.220 & 5.34 $\pm$ 0.31 &  2.42 & 18.60 & 33.04  \\ 
  18 & 5712.25 $\pm$ 1.18 & 1.170 $\pm$ 0.390 & 7.29 $\pm$ 2.83 &  0.40 &  3.00 &  4.86  \\ 
  19 & 6330.57 $\pm$ 0.59 & 1.600 $\pm$ 0.380 & 4.20 $\pm$ 0.00 &  0.44 &  4.20 &  4.55  \\ 
  20 & 6343.12 $\pm$ 0.48 & 1.670 $\pm$ 0.320 & 4.20 $\pm$ 0.00 &  1.77 &  5.30 &  5.16  \\ 
  21 & 6585.49 $\pm$ 0.88 & 1.120 $\pm$ 0.380 & 5.39 $\pm$ 2.10 &  0.09 &  3.00 &  4.01  \\ 
\tablenotetext{a}{Wavelengths are vacuum heliocentric.} 
\tablenotetext{b}{Lines with a value of $4.20 \pm 0.00$ were fit with the FWHM set to the minimum allowed value.} 
\tablenotetext{c}{Significance of the line defined as $EW/\sigma_{fit}$, where $\sigma_{fit}$ is the error in the measured equivalent width.} 
\tablenotetext{d}{Significance of the line defined as $EW/\sigma_{det}$, where $\sigma_{det}$ is the $1\sigma$ limiting equivalent width.} 
\tablenotetext{e}{This line has been identified with an associated absorption
system and is included in Table 3.}
\enddata 
\end{deluxetable}

% Table 3
\begin{deluxetable}{rccccrc} 
   
\tablenum{3} 
\tablewidth{0pt}
\tablecaption{Associated Absorption} 
\tablehead{ 
 \multicolumn{1}{c}{}  & 
 \multicolumn{2}{c}{Component A}  & 
 \multicolumn{2}{c}{Component B}  & 
 \multicolumn{1}{c}{}  & 
 \multicolumn{1}{c}{}  \\ 
 \cline{2-3} \cline{4-5} \\
 \colhead{Line}  & 
 \colhead{Central $\lambda$}  & 
 \colhead{Redshift}  & 
 \colhead{Central $\lambda$}  & 
 \colhead{Redshift}  & 
 \colhead{Identification}  & 
 \colhead{$\Delta$ v}  \\ 
 \colhead{}  & 
 \colhead{(\AA)}  & 
 \colhead{}  & 
 \colhead{(\AA)}  & 
 \colhead{}  & 
 \colhead{}  & 
 \colhead{(km s$^{-1}$)}  
}   
\startdata 
   1 & 4228.10 $\pm$ 0.02 & 2.4780 &  4221.22 $\pm$ 0.58 & 2.4723 & \ion{H}{1} 1216    &  488.2 $\pm$ 41.2    \\
   2 & 4306.59 $\pm$ 0.12 & 2.4764 &  4301.56 $\pm$ 0.26 & 2.4723 & \ion{N}{5} 1238   &   350.3 $\pm$ 20.0  \\
   3 & \nodata\tablenotemark{a} & \nodata &  4315.10 $\pm$ 0.51 & 2.4721 & \ion{N}{5} 1242  & \nodata \\
   4 & 4846.76 $\pm$ 0.67  & 2.4775 &  4839.88 $\pm$ 0.62 & 2.4726 & \ion{Si}{4} 1393  &   425.9 $\pm$ 56.5  \\
   5 & \nodata\tablenotemark{b} & \nodata &  4870.19 $\pm$ 0.35  & 2.4718 & \ion{Si}{4} 1402  & \nodata \\
   6 & 5384.81 $\pm$ 0.06 & 2.4811 &  5375.40 $\pm$ 0.13 & 2.4720 & \ion{C}{4} 1548   &   524.3 $\pm$ 8.0  \\
   7 & 5392.90 $\pm$ 0.08 & 2.4776 &  5384.23 $\pm$ 0.13 & 2.4720 & \ion{C}{4} 1550   &   482.3 $\pm$ 8.5  \\

\enddata 
\tablenotetext{a}{
Both components of the \ion{N}{5} doublet are detected in absorption in 
Component B.  However, because \ion{N}{5} emission is much stronger in 
Component A, ambiguities in continuum fitting have made it difficult to 
detect \ion{N}{5} 1242 in absorption for this object.}
\tablenotetext{b}{
This absorption feature was detected in the spectrum of the A component,
but it did not meet the significance criteria and is not included in 
Table 1.}

\end{deluxetable}

\end{document}